\documentclass[apj]{emulateapj}
\shorttitle{Twist, Writhe, and Helicity}
\shortauthors{B. Ruiz Cobo \& K.~G. Puschmann}

\usepackage{hyperref}
\hypersetup{
     final=true,
     pageanchor=true,
     hyperfigures=false,
     colorlinks=true,
     breaklinks=true,
     linkcolor=blue,
     citecolor=blue,
     urlcolor=blue,
     pdfpagemode=UseNone,
     pdfpagelayout=Two Column,
     pdfview=FitBH,
     pdftitle={Twist, Writhe and Helicity in the Inner Penumbra of a Sunspot},
     pdfauthor={B. Ruiz Cobo, K.G. Puschmann},
     pdfsubject={Solar Physics}}

\begin{document}


\title{Twist, Writhe, and Helicity in the inner penumbra of a Sunspot}


%% \author{K. G. Puschmann\altaffilmark{1,2,3}, B. Ruiz Cobo\altaffilmark{2,3} and V. Mart\'\i nez Pillet\altaffilmark{2}}
%% 
%% \affil{(1) Astrophysikalisches Institut Potsdam (AIP), D-14482, Potsdam, Germany\\
%% (2) Instituto de Astrof\'isica de Canarias (IAC), E-38200 La Laguna, Tenerife, Spain\\
%% (3) Departamento de Astrof\'isica, Universidad de La Laguna (ULL), E-38205 La Laguna, Tenerife, Spain}
%% \email{kgp@aip.de, brc@iac.es, vmp@.iac.es}

\author{B. Ruiz Cobo\altaffilmark{1,2,3}  and K. G. Puschmann\altaffilmark{4,1,3}}

\affil{
(1) Instituto de Astrof\'isica de Canarias (IAC), E-38200 La Laguna, Tenerife, Spain\\
(2) National Solar Observatory (NSO), 950 North Cherry Avenue, Tucson, AZ 85719, USA\\
(3) Departamento de Astrof\'isica, Universidad de La Laguna (ULL), E-38205 La Laguna, Tenerife, Spain\\
(4) Leibniz-Institut f\"ur Astrophysik Potsdam (AIP), D-14482, Potsdam, Germany}
\email{brc@iac.es, kgp@aip.de}


%\altaffiltext{1}{Visiting Astronomer, Cerro Tololo Inter-American Observatory.
%CTIO is operated by AURA, Inc.\ under contract to the National Science
%\altaffiltext{2}{Society of Fellows, Harvard University.}
%\altaffiltext{3}{present address: Center for Astrophysics,
%    60 Garden Street, Cambridge, MA 02138}
%\altaffiltext{4}{Visiting Programmer, Space Telescope Science Institute}
%\altaffiltext{5}{Patron, Alonso's Bar and Grill}

\begin{abstract}
The aim of this work is the determination of the twist, writhe, and self magnetic 
helicity of penumbral filaments located in an inner Sunspot penumbra. To this extent, 
we inverted data taken with the spectropolarimeter (SP) aboard {\it Hinode} with the 
SIR (Stokes Inversion based on Response function) code. For the construction 
of a 3D geometrical model we applied a genetic algorithm minimizing the divergence
of $\vec {B}$ and the net magnetohydrodynamic force, consequently a force-free solution
would be reached if possible.
 We estimated two proxies 
to the magnetic helicity frequently used in literature: the force-free parameter 
$\alpha_z$ and the current helicity term $h_{c_{z}}$. We show that both proxies are 
only qualitative indicators of the local twist as the magnetic field in the area 
under study significantly departures from a force-free configuration. The local 
twist shows significant values only at the borders of bright penumbral filaments 
with opposite signs on each side. These locations are precisely correlated to large 
electric currents. The average twist (and writhe) of penumbral structures 
is very small. The spines (dark filaments in the background) show a nearly zero 
writhe. The writhe per unit length of the intraspines diminishes with increasing 
length of the tube axes. Thus, the axes of tubes related to intraspines are less 
wrung when the tubes are more horizontal. As the writhe of the spines is very 
small, we can conclude that the writhe reaches only significant values when the 
tube includes the border of a intraspine.

\end{abstract}


\section{Introduction}
\label{intro}
Investigating the physical nature and the dynamics of penumbral filaments is essential in order to understand the 
structure and the evolution of sunspots and their surrounding moat regions. Many important observational 
aspects of penumbral filaments are well settled down, although their interpretation is still a source of 
debate, e.g., the brightness of penumbral filaments, the inward motion of bright penumbral grains, the 
Evershed flow, the Net Circular Polarization (NCP), as well as moving magnetic features in the sunspot moat. 
During the last few years, new observational discoveries, e.g., dark cored penumbral filaments 
\citep{scharmeretal02}, strong downflow patches in the mid and outer penumbra \citep{ichimotoetal07a}, 
penumbral micro-jets \citep{katsukawaetal07, jrcakkatsu08}, or twisting motions of penumbral filaments 
\citep{scharmeretal02,rimmelemarino06,ichimotoetal07b,ningetal09} broadened the number of unknowns and 
gave new impulse to the investigation of sunspots. For recent reviews see \citet{borrero11}, \citet{bellot10}, \citet{borrero09}, 
\citet{sch09}, or \citet{tritschler09}. An especially controversial issue is the study of the twisting 
motions of penumbral filaments. On the one hand, \citet{ichimotoetal07b} consider twisting motions as an 
apparent phenomenon, produced by lateral motions of intensity fluctuations associated with overturning convection. 
On the other hand, \citet{ryutovaetal08} propose that the observed twist is an intrinsic property of penumbral 
filaments and is produced as a consequence of reconnection processes which take place in the penumbra. 
\citet{suetal08,suetal10} conclude that the twist of penumbral filaments changes with time 
caused by an unwinding process. 

In the present paper we study the twist of filaments of the inner penumbra of a sunspot by means of the magnetic 
helicity. We take advantage of the 3D geometrical model of a section of the inner penumbra of a sunspot described 
in \citet{puschmannetal10a} [hereafter, Paper I]. We use observations of the active region AR 10953 near solar disk 
center obtained on 1$^{\rm st}$ of May 2007 with the  Hinode/SP. The inner, center side, penumbral area under study 
was located at an heliocentric angle $\theta$\,=\,4.63$^{\circ}$. To derive the physical parameters of the solar 
atmosphere as a function of continuum optical depth, the SIR (Stokes Inversion based on Response function) 
inversion code \citep{ruizcobodeltoro92} was applied to the data set. The 3D geometrical model was derived by 
means of a genetic algorithm that minimized the divergence of the magnetic field vector and the deviations from 
static equilibrium considering pressure gradients, gravity and the Lorentz force. 
We can not assess the unicity of the resulting model: the found solution just minimizes the divergency of 
the magnetic field and the modulus of the net force, neglecting the contribution of the aceleration terms.
For a detailed description we refer to Paper I. In \citet{puschmannetal10b} [hereafter, Paper II], we calculated 
the electrical current density vector $\vec{J}$ in the above mentioned area and found the horizontal component 
of the electrical currents $\sim 4$ times larger than the vertical component ${J}_z$ \citep[thus confirming the 
results of][]{pevtsovandperegud90, georgoulislabonte04}. In addition, we concluded that the magnetic field at the 
borders of bright penumbral filaments departs from a force-free configuration \citep[see also][]{zhang10}.
These results are strongly significant considering that we have imposed that our solution minimizes the net force,
including the Lorentz force, and  consequently, a force free solution should be found if it were possible.
\begin{figure}
   \begin{center}
      \includegraphics[scale=0.6]{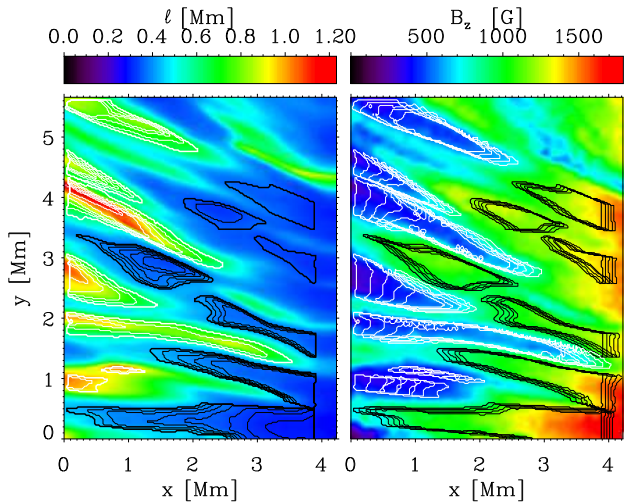}
    \end{center}
    \caption{Left panel: Length $\ell$ of the field lines of  $\vec{B}$ integrated from the top 
($z=200$\,km) to the bottom layer ($z=0$\,km) or until $x=4.2$\,Mm. 
Thick isolines of equal $\ell$ define 14 areas (flux tubes) related to spines (black) and intraspines (white), 
respectively.
Thin isolines denote different thresholds for selecting the flux tube area.
 Right panel: Vertical component of $\vec{B}$ at $z$\,=\,200\,km. 
 Contour lines correspond to horizontal cuts through the 14 flux tubes with the largest area 
at $z=200, 150, 100, 50, 0$\,km (the thickness of each line diminishes with depth). Black and white contours 
distinguish between flux tubes related to spines and intraspines, respectively.}
    \label{Fig1}
\end{figure}

We can evaluate the magnetic field lines by the integration of $\vec{B}$ starting at each pixel
of the top layer of our volume of the inner penumbra (of 4.2 Mm $\times$ 5.6 Mm $\times$ 
0.2 Mm, see Paper II). In the left panel of Fig.~\ref{Fig1} we present the length $\ell$ 
of each field line. As our sunspot has negative polarity, $\vec{B}$ points downwards and 
we thus integrate the field lines from the top layer to the bottom layer.  The majority of field 
lines end up at the bottom layer, except the lines starting at larger X coordinates 
in our FOV.  
In the whole analized volume the magnetic field has the same polarity, and consequently,
the field lines do not present maxima nor minima in our region, i.e., the field lines always travel
 downwards. This fact, as we will see later, simplifies the evaluation of the magnetic helicity.
Areas with larger $\ell$ correspond to intraspines, since $\vec{B}$ is more horizontal, the field 
lines thus traverse larger distances inside our volume.

In the left panel of Fig.~\ref{Fig1}, we selected 14 areas, 7 corresponding to intraspines (thick white 
contours) and  7 to spines (thick black contours), according to the length $\ell$ of the magnetic field lines. 
For each of the selected zones we define a volume delimited by the field lines setting off from each 
pixel of the closed curve of the top layer. 
In the right panel of Fig.~\ref{Fig1} we show the vertical component of $\vec B$ at 
$z$\,=\,200\,km together with horizontal cuts through each of these volumes at 
$z=200, 150, 100, 50, \&\,0$\,km. Since the umbra is placed at the right hand side in 
the FOV, the cuts at deeper layers are displaced to the right. Finally we checked that the 
magnetic flux traversing each of the cuts is approximately constant: 
the standard deviation of the relative variation of the magnetic flux between the top
and the bottom layer is 4.5\%. Thus, each volume can be approximately 
considered as a flux tube.

However, the 14 areas were selected quit arbitrarily. In order to study the dependence
of twist, writhe and magnetic helicity on the flux tube area, 40 additionally smaller 
tubes have been defined inside the larger 
ones (thin isolines in the left panel of Fig.~\ref{Fig1}). Thus, for most 
of the 14 zones we have several tubes of different size, many of 
them being the internal part of the larger
one. Consequently, we have 54 tubes, 27 of them 
related to intraspines and 27 to spines.

\section{Magnetic helicity, Twist, and Writhe}
The study of helicity of solar magnetic features has been a hot topic during at least the last 25 years. 
Magnetic helicity has been investigated in solar structures at different spatial scales in the 
photosphere and chromosphere, as well as in the solar wind \citep[see e.g. the reviews of]
[and references therein]{brownetal99, rust02, pevtsovandbala03, demoulin07, demoulinandpariat09}. 
The helicity in penumbral filaments has been analyzed by means of some proxies by, e.g., 
\citet{ryutovaetal08}, \citet{tiwariandvenkata09}, \citet{suetal10}, and \citet{zhang10}.
\begin{figure*}
  \begin{center}   
    \includegraphics[scale=1.3]{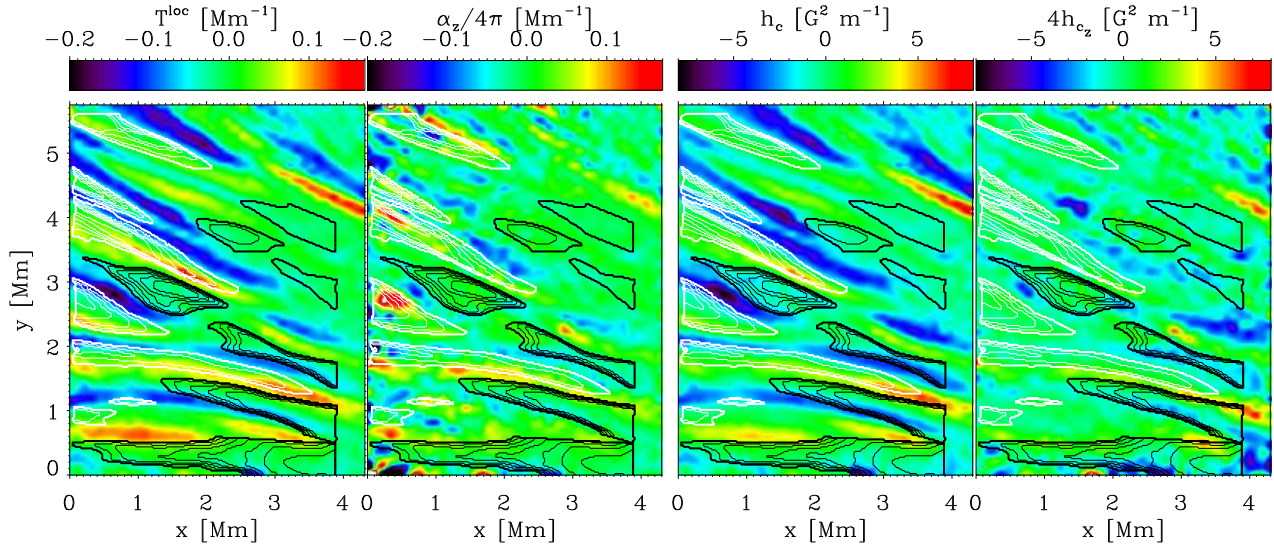} 
  \end{center}
   \caption{From left to right: local twist ($T^{\mbox{\rm\tiny{loc}}}$), $\alpha_{z}/4\pi$, 
current helicity density ($h_{c}$), and $4h_{c_{z}}$ evaluated at $z=200$\,km. 
The isolines are the same as in the left panel of Fig.~\ref{Fig1}.}
   \label{Fig2}
\end{figure*}

The magnetic helicity, $H_{m}$, quantifies how the magnetic field is twisted,  
writhed, and linked.
$H_{m}$ plays a key role in magneto-hydrodynamics because 
it is almost conserved in a plasma having a high magnetic Reynolds number \citep[see e.g.,][]
{berger84}. The magnetic helicity of a vector field $\vec{B}$, fully contained within a volume 
$\mathcal{V}$ and bounded by a surface $\mathcal{S}$ (i.e., the normal component 
$B_{n}=\vec{B}\cdot\vec{n}$ vanishes at any point of $\mathcal{S}$), is \citep{elsasser56}:
\begin{equation}
H_{m}=\int_{\mathcal{V}}\vec{A}\cdot\vec{B}\,\, \mathrm{d}^{3}x,
\label{eq1}
\end{equation}
where the vector potential $\vec{A}$ satisfies $\vec{B}=\nabla \times \vec{A}$. 
\citet{bergerandfield84} showed that Eq.~\ref{eq1}, 
is not gauge-invariant if the volume of interest is not bounded by a magnetic surface, i.e., if $\vec{B}$ 
crosses $\mathcal{S}$ (as in the case of the volume of the penumbra retrieved from our observations). 
In this case the relative magnetic helicity \citep{finnandandantonsen85} should be used: 
\begin{equation}
H_{m}^{rel}=\int_{\mathcal{V}}(\vec{A}+\vec{A}_{p})\cdot(\vec{B}-\vec{B}_{p})\,\, \mathrm{d}^{3}x,
 \label{relativehm}
\end{equation}
where $\vec{B}_{p}$ is a potential field having the same normal component $B_{n}$ on $\mathcal{S}$, 
and $\vec{A}_{p}$ is its vector potential. 
The relative helicity reflects twist, writhe, and linkage with respect to a current-free (potential) field,
i.e., its minimum-energy state for the given $B_{n}$-condition on $\mathcal{S}$. 
The relative magnetic helicity so defined is gauge-invariant 
and has the same conservation properties and amount of topological information 
as the magnetic helicity. 
Throughout this article, the term magnetic helicity refers to the relative magnetic helicity. For an isolated
magnetic flux rope, $H_{m}^{rel}$ is proportional to the sum of its twist $T$ and writhe $W$ 
\citep[][]{bergerandfield84, Toroketal10}:
\begin{equation}
H_{m}^{rel}=(T+W)\Phi^{2}
 \label{TandW}
\end{equation}
where $\Phi$ is the magnetic flux of the rope. The twist is the turning angle of a bundle of 
magnetic field lines around its central axis, whereas the writhe quantifies the helical deformation 
of the axis itself. Following \citet{bergerandprior06}, the twist of an infinitesimal rope is given by 
$T=\int T^{\mbox{\rm\tiny{loc}}}{\mathrm{d}l}$, $l$ being the arc length along the central  
field line of the rope and $T^{\mbox{\rm\tiny{loc}}}$ the local twist:
\begin{equation}
T^{\mbox{\rm\tiny{loc}}}=\frac{\mathrm{d}T}{\mathrm{d}l}=\frac{\mu_{0}J_{||}}{4 \pi B_{||}}\,\,,
 \label{local}
\end{equation}
being $J_{||}$ and $B_{||}$ the components of the current and magnetic field parallel to the central  
field line of the rope. With this definition, $T=1$ when the field lines twist around the axis by an 
angle of $2\pi$ and $T^{\mbox{\rm\tiny{loc}}}$ is the local twist per unit length evaluated at a given 
geometrical height at each pixel.
If the rope has a non-infinitesimal cross section $\Sigma$, the local twist of the rope 
$T^{\mbox{\rm\tiny{loc}}}_i$ is given by the average of the infinitesimal local twist, 
$T^{\mbox{\rm\tiny{loc}}}$ over $\Sigma$.
\citet[][see also \citet{Toroketal10}]{bergerandprior06} give expressions for the writhe of specific 
geometrical configurations.
Provided that the magnetic field lines in the inner penumbral region studied in this paper always
travel downwards in $z$, i.e. without showing maxima nor minima between the $z_0=$200\,km and $z_1=$0\,km 
height layers, we can evaluate the writhe using a very simplified formula: 
\begin{equation}
 W=\frac{1}{2 \pi}\int_{z_0}^{z_1}\frac{1}{1+|\tau_z|}{(\vec{\tau} \times \vec{\tau'})}_{z}{\mathrm{d}z}\,,
 \label{writhe}
\end{equation}
obtained as a particular case of the more general formula of \citet{bergerandprior06}.
In Eq.~\ref{writhe}, $\vec{\tau}$ stands for the tangent vector to the tube axis;  
${\tau_z}$ for the vertical component of $\vec{\tau}$; and 
$\vec{\tau'}=\frac{\mathrm{d}\vec{\tau}}{\mathrm{d}z}$.

Equations~\ref{TandW}, \ref{local}, \& \ref{writhe} allow the evaluation of the twist, 
the writhe and the magnetic helicity of an isolated tube. 
What happens in the case of a non-isolated tube, as is clearly the case of the penumbral tubes?
The writhe of a magnetic rope, isolated or not, is a measure of the helical deformation of
the axis of the rope, while the twist quantifies the winding of the magnetic field lines
of the rope around its axis. The magnetic helicity measures the linking number of the
field lines, averaged over all pairs of lines, and weighted by the flux \citep{bergerandprior06, Moffatt69}.
 We can simplify the case of a non-isolated tube to a scenario
in which we have just two adjoining tubes. It is clear that we can define the writhe and
twist for each individual tube, and consequently its magnetic helictiy, but the helicity of the 
whole configuration is not just the sum of both contributions. We would need to include an extra
term taking into account the linking between both tubes: the mutual helicity.
Consequently the application of equations~\ref{TandW}, \ref{local}, \& \ref{writhe}
to our tubes retrieves only the contribution of the local values of the twist, writhe and the 
self helicity. The contribution of the surrounding tubes to the helicity 
of each flux tube is not considered. This contribution, the mutual helicity, could 
be larger than the self helicity \citep[see e.g.,][]{renierpriest07}. 
 The mutual helicity can be calculated using the procedure described 
in \citet{bergerandprior06}. However, for the scope of this paper we limit the calculation 
to self helicity.

\section{Proxies of the magnetic helicity}
 The current helicity density is defined \citep[see e.g.,][]{seehafer90} as 
$h_{c}={\vec B}\cdot\nabla\times{\vec B}$. 
 If we take as magnetic ropes the tubes defined by the field lines starting in each pixel, 
(being then $B_{||}=B$), Eq.~\ref{local} becomes $T^{\mbox{\rm\tiny{loc}}} = h_{c} / 4 \pi B^{2}$. 
The parameter $\alpha$ is usually defined in a force-free configuration, i.e., when  ${\vec B}$ is 
parallel to its curl, by $\nabla\times{\vec B}$\,=\,$\alpha {\vec B}$.
The $\alpha$ parameter can be defined
for non force-free fields: $\alpha= {\vec B}\cdot\nabla\times{\vec B}/B^{2}$. This is the 
definition\begin{footnote}{\citet{yeatesetal08} denominate current helicity to this generalized 
$\alpha$ parameter.}\end{footnote} we will use throughout the paper. 
In order to see how this re-defined
$\alpha$ differs from the force-free definition, let us decompose 
$\nabla\times{\vec B}=(\nabla\times{\vec B})_{||} +  (\nabla\times{\vec B})_{\perp}$.
From its definition,  
$\alpha$ becomes equal to 
$\alpha=\pm|(\nabla\times{\vec B})_{||}|/B$ which is equal to the classical 
definition for a force-free field. The $\pm$ is needed to consider the case 
when $(\nabla\times{\vec B})_{||}$ and $\vec B$ point in opposite direction, i.e., 
when $\alpha$ is negative.
For a no force free-field, the parameter $\alpha$ is then
the ratio between the parallel component of the curl of the magnetic field and its modulus.
Evidently, in a general case, we will
have $\alpha=h_{c}/B^{2}=4\pi\,T^{\mbox{\rm\tiny{loc}}}$. 

Given the difficulty of empirically obtaining $H_{m}^{rel}$ and ${\vec J}$, one finds many works where 
different proxies were used. Before evaluating the magnetic helicity we can calculate, from 
our data, some of the most usual proxies of the magnetic helicity. Among them the most common 
proxies, with several different but more or less equivalent definitions, are the 
$\alpha_{z}=(\nabla\times{\vec B})_{z}/B_{z}$ parameter 
\citep[see e.g.,][and references therein]{suetal09, suetal10, pevtsovetal08}, and the 
parameter $h_{c_{z}}=B_{z}(\nabla\times{\vec B})_{z}$ \citep[see e.g.,][]{zhang10}. It is evident 
that for a force-free field $\alpha_{z}=\alpha$ and $h_{c_{z}}=\alpha B_{z}^2=h_{c}B_{z}^2/B^2$. 
That means that 
the $h_{c_{z}}$ parameter could be meaningless for nearly horizontal magnetic fields, 
such as those found in sunspot penumbrae.

Many authors suppose that  
the sign of the integral of $\alpha_{z}$ (or $h_{c_{z}}$) over the volume of a magnetic structure 
coincides with the sign of $H_{m}^{rel}$, although this fact has not yet been demonstrated 
\citep{demoulin07}. On the other hand, \citet{hagyardpevtsov99} point out that $h_{c_{z}}$ only 
considers the vertical component $J_{z}$ of the 
electric current density vector, $h_{c_{z}}$ can 
strongly differ from $h_{c}$, provided that $J_{z}$ is much smaller than the horizontal 
components, at least in the inner penumbra (see Paper II). Besides, \citet{pariatetal05} comment 
that, since the magnetic helicity is a global quantity, it is not obvious that a helicity density 
has any physical meaning.

We evaluated these proxies in the inner penumbral region under study.
In the left panels of Fig.~\ref{Fig2} we present $T^{\mbox{\rm\tiny{loc}}}$ (evaluated from Eq.~\ref{local}) 
and $\alpha_{z}/4\pi$ evaluated at each pixel at $z=200$\,km. As in the 
inner penumbra the magnetic field is not force-free at the borders of bright penumbral filaments 
(see Paper II), $\alpha_{z}/4\pi$ (2$^{\rm nd}$ panel) is only qualitatively similar to 
$T^{\mbox{\rm\tiny{loc}}}$. In the 3$^{\rm rd}$ and 4$^{\rm th}$ panel we present $h_{c}={\vec B}\cdot\nabla\times{\vec B}$
and $h_{c_{z}}$ evaluated at $z=200$\,km. $h_{c_{z}}$ is multiplied by a factor 4 just to make 
easier its comparison with $h_{c}$. As we have seen before, 
$T^{\mbox{\rm\tiny{loc}}} = h_{c} / 4 \pi B^{2}$, and thus the general aspect of $h_{c}$ is 
very similar to  $T^{\mbox{\rm\tiny{loc}}}$. However, $h_{c_{z}}$ 
resembles $h_{c}$ only marginally (the standard deviations are 
$\sigma(h_{c})=2.1\,G^{2}\,m^{-1}$ and $\sigma(h_{c_{z}})=0.4\,G^{2}\,m^{-1}$). 
The values of $\alpha_{z}/4\pi$ and $h_{c_{z}}$ at $z=200$\,km obtained here are 
very similar to the results found in the literature: \citet{tiwariandvenkata09} found 
that $\alpha_{z}/4\pi$ varies around $\pm 0.15 Mm^{-1}$ along azimuthal paths in the 
middle penumbra; \citet{suetal10} found a fluctuation of $\alpha_{z}/4\pi$ larger than 
$\pm 0.05 Mm^{-1}$ over an inner penumbral region; \citet{suetal09} found that $h_{c_{z}}$ 
fluctuates along an azimuthal path in the inner penumbra with an amplitude larger than 
1\,$G^{2}\,m^{-1}$ while \citet{horstpeter08} found penumbral mean values of about 
0.04\,$G^{2}\,m^{-1}$. In the four panels of Fig.~\ref{Fig2}, the outlined areas were 
selected by different thresholds of $\ell$, the length of the magnetic field lines
between the layers $z=200$\,km and $z=0$\,km (see also Section~\ref{intro} and Fig.~\ref{Fig1}).
The sign of the integral of $T^{\mbox{\rm\tiny{loc}}}$ and $\alpha_{z}/4\pi$ over 
the above mentioned areas only coincides in 39\% of the 54 tubes (if we consider only the areas related 
to the intraspines this value decreases to 15\% of the 27 tubes). The same figures are obtained for the
percentage of coincidence between the signs of the integrals of $h_{c}$ and $h_{c_{z}}$, approximately.
This weak coincidence demonstrates that, at least in the inner penumbra of a sunspot, 
$\alpha_{z}/4\pi$ and $h_{c_{z}}$ are not good estimates of 
$T^{\mbox{\rm\tiny{loc}}}$ and $h_{c}$, respectively.  As already shown in Paper II, the magnetic field in 
the area under study significantly departures from a force-free configuration. 

Note that $T^{\mbox{\rm\tiny{loc}}}$ 
reaches significant values only at the borders of the intraspines: these are exactly the 
areas where the electric current density is large (see Fig. 1 of Paper II). 
Furthermore, often $T^{\mbox{\rm\tiny{loc}}}$ changes its sign at both sides of bright 
filaments, i.e., for the majority of the intraspines, $T^{\mbox{\rm\tiny{loc}}}$ shows 
negative values at the upper (larger Y-coordinate) borders of the filaments and positive 
values at the lower borders. The alternation of the sign in the twist is also observed 
(although less evident) in the maps of $\alpha_{z}$ and $h_{c_{z}}$ and it is clearly 
visible in \citet{tiwariandvenkata09}, \citet{suetal09}, \citet{suetal10}, and \citet{zhang10}.

This phenomenon can be explained if we consider that the field lines of the magnetic background 
component wrap around the intraspines \citep{borreroetal08} and tend to meet above the 
intraspines, thus generating a curvature of different sign in the field lines at both 
sides of the intraspines. Thus $T^{\mbox{\rm\tiny{loc}}}$ could be measuring the twist of 
the background field wrapping around the intraspines rather than the twist of the field lines 
of the intraspines themselves. However, the alternation of signs of the twist at both sides 
of a penumbral filament is compatible with the magnetohydrostatic equilibrium model of a 
magnetic flux tube built by \citet{borrero07}. This model includes a transverse component 
of $\vec{B}$ having opposite twist at both sides of a plane longitudinally cutting the flux 
tube. This model is able of explaining both the dark cored penumbral filaments and the net 
circular polarization observed in penumbral filaments \citep{borreroetal07}. \citet{magara10} 
suggests the existence of an intermediate region where the magnetic field has a 
transitional configuration between a penumbral flux tube and the background field: in such areas, 
coinciding with the largest electrical current density (see Paper II), penumbral micro-jets 
are produced as observed by \citet{katsukawaetal07}.

\section{Numerical test}
To check its correctness, the procedure used for the evaluation of twist, writhe, and magnetic 
helicity was applied to two different analytical cases. 
In the first case we consider that the magnetic field lines follow a helix around a vertical 
straight line. The magnetic field vector is defined by 
$\vec{B}=B_0\,\hat{z} + B_1\,r\,\hat{\theta} $
with $B_0$ and $B_1$ being constant. 
$\vec{B}$ could easily be decomposed in a potential $\vec{B}_p=B_0\,\hat{z}$ 
and a close (toroidal) field $\vec{B}_c=B_1\,r\,\hat{\theta}$. 
Obviously, the potential component $\vec{B}_p$ fulfills the conditions required 
in Eq.~\ref{relativehm} in the case of a cylindrical tube: $\vec{B}$ and $\vec{B}_{p}$ 
have the same normal component on the external surface of the tube. 
The determination of the vector potential, in this case, is straightforward: 
$\vec{A}=B_0\,r/2\,\hat{\theta} - B_1\,r^2/2\,\hat{z}$. 
The vector potential of the potential component will be 
$\vec{A_p}=B_0\,r/2\,\hat{\theta}$. 
Using Eq.~\ref{relativehm}, the magnetic helicity of a cylindrical
tube of height $L$ and radius $R$ becomes 
$H_m=\frac{1}{2}\pi\,B_0\,B_1\,R^4\,L$.
The magnetic flux of this tube is $\Phi=\pi\,B_0\,R^2$. 
This tube has a zero writhe because its axis is a straight line.  
From Eq.~\ref{TandW}, the twist follows as 
\begin{equation}
  T=\frac{B_1\,L}{2\pi\,B_0}.
 \label{Analitic_Twist}
\end{equation}
This is obviously the expected result, provided that the pitch (of screw-step) of our helix is 
$\frac{2\pi\,B_0}{B_1}$ and the twist is a measure of the number of turns done by 
the magnetic field lines along a longitude $L$.
 
In the first four rows of Table \ref{table1} we present the analytical 
(i.e., using Eq.~\ref{Analitic_Twist}) and numerical results (using Eqs.~\ref{local} 
and \ref{writhe}) for tubes with $B_0=-0.2\, {\mbox{T}}$, $B_1=0.04\, 
{\mbox{T}}\,{\mbox{Mm}}^{-1}$, $L=0.225\,{\mbox{Mm}}$ and a radius $R$ equal to $0.2$ 
and $1$ Mm. We used the same spatial grid as in the observational case.

In order to check the accuracy of the determination of twist, writhe and magnetic helicity  in
a more general case, we carried out a second test, building a helical magnetic field that
turns around a helical axis. Let us suppose that the axis of the tube is a vertical
helix of radius $R_h$ turning an angle $\Psi$ through a length $L=0.22\,{\mbox{Mm}}$. Then, the 
magnetic field at the axis will be $\vec{B}=B_0\, \hat{z} + B_1\,R_h\, \hat{\theta} $ with
$B_1=B_0\,\Psi/L$. Following \citet{bergerandprior06}, the writhe of a magnetic flux tube 
whose axis is a helix can be easily evaluated in terms of its polar writhe (i.e., area/$2\pi$ of  
the section of the unity sphere limited by the tantrix curve and the north pole. 
The tantrix curve is the path, the tip of the tangent vector takes on the unit sphere). 
In our case, the writhe becomes: 
\begin{equation}
  W=\frac{\Psi}{2\pi}(1-\frac{B_0}{|B|}).
 \label{Analitic_Writhe}
\end{equation}
Once we have the axis, we can easily build a tube with a given twist
around such an axis. We chose the radius of the tube as $R=0.2$ Mm, and the radius of 
the helical axis as $R_h=0.4\,{\mbox{Mm}}$, and $B_1=0.5\,{\mbox{T\,Mm$^{-1}$}}$. The angle $\Psi$ 
takes a value of $9.124$ degrees in order to have an analytical writhe 
(i.e., using Eq.~\ref{Analitic_Writhe}) of $0.001$. 
The added twist takes the values $0.0$, $0.001$, $0.01$, $0.1$, and $-0.001$. 
Note that, as the magnetic helicity is proportional to the sum of twist and writhe, in 
the last case we will have a null magnetic helicity. The results of these tests are 
presented in the ten bottom rows of Table \ref{table1}.  

\begin{table}
  \caption{Test results: Writhe ($W$), twist ($T$), and magnetic helicity ($H_m$), for a helical 
magnetic field (first four rows) and a helical magnetic field winding around a helical axis.}
  \label{table1}
  \begin{center}
    \leavevmode
    \begin{tabular}{ccccc} \hline \hline              
  $R$ [Mm]       &           & $W$     &  $T$     &  $H_m\,\,[{\mbox {Mx}}^2]$  \\ \hline 
  0.2            & analytical&  0.00   & -7.16e-3 &    -4.52e+34         \\ 
                 & numerical & -9.e-28 & -7.16e-3 &    -4.50e+34         \\ \hline
  1.0            & analytical&  0.00   & -7.16e-3 &    -2.83e+37         \\ 
                 & numerical & -9.e-28 & -7.16e-3 &    -2.81e+37         \\ \hline
  0.2            & analytical& 1.00e-3 &  0.00    &     8.16e+34          \\ 
                 & numerical & 0.99e-3 &  1.90e-6 &     8.07e+34          \\ \hline
  0.2            & analytical& 1.00e-3 &  1.00e-3 &     1.63e+35          \\
                 & numerical & 0.99e-3 &  1.07e-3 &     1.67e+35          \\ \hline
  0.2            & analytical& 1.00e-3 &  1.00e-2 &     8.97e+35          \\
                 & numerical & 0.98e-3 &  1.06e-2 &     9.47e+35          \\ \hline
  0.2            & analytical& 1.00e-3 &  1.00e-1 &     7.01e+36          \\
                 & numerical & 0.97e-3 &  0.99e-1 &     6.91e+36          \\ \hline
  0.2            & analytical& 1.00e-3 & -1.00e-3 &     0.00             \\
                 & numerical & 0.99e-3 & -1.06e-3 &    -6.14e+33          \\ \hline
    \end{tabular}
  \end{center}
\end{table}

\begin{figure}
   \begin{center}
      \includegraphics[scale=0.5]{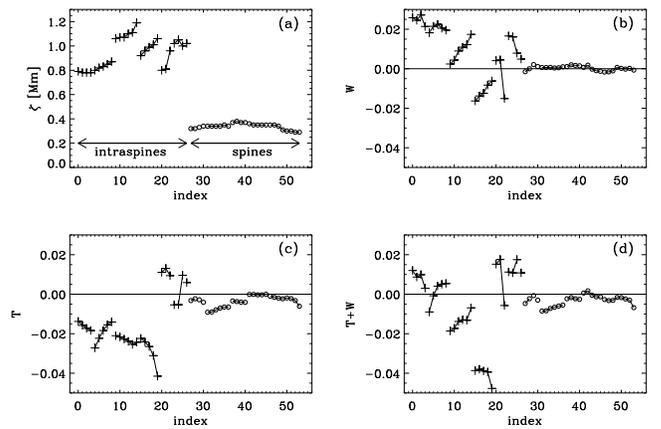} 
   \end{center}
   \caption{Panel (a): Length of the axis of each flux tube. The values 
corresponding to tubes of the same zone are connected by lines. Cross symbols correspond to intraspines and small circles to spines.
Panels (b), (c), and (d): writhe, twist, sum of writhe and twist.}
   \label{Fig3}
\end{figure}
\begin{figure}
   \begin{center}
      \includegraphics[scale=.50]{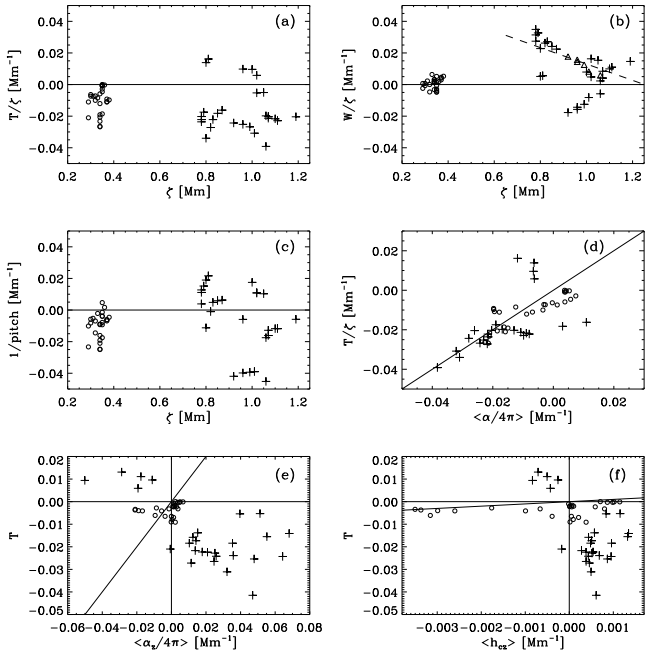} 
   \end{center}
   \caption{Panels (a) and (b): normalized twist ($T/\zeta$) and writhe ($W/\zeta$) {\it versus}
 the length of the axis of each flux tube $\zeta$. The absolute value $|W/\zeta|$ has been 
overplotted in panel (b) with triangle symbols. The dashed line is the linear fit of 
$|W/\zeta|$. 
Panel (c): wavenumber. 
Panel (d): $T/\zeta$ versus $\alpha/4\pi$ averaged over the section of each flux tube at $z=200$\,km. 
Panels (e) and (f): $T$ versus the average of $\alpha_z/4\pi$ and $h_{c_{z}}$
respectively.  In panels (d), (e) and (f) the straight line with slope 1 has been overplotted. 
Cross symbols correspond to intraspines and small circles to spines.
}
   \label{Fig4}
\end{figure}

\section{Results}
\begin{table}
  \caption{Axis length $\zeta$, magnetic flux $\Phi$, twist, writhe and 
magnetic helicity for the 14 largest zones. The 7 first (last) rows are related to 
intraspines (spines).} 
  \label{table2}
  \begin{center}
    \leavevmode
    \begin{tabular}{cccccc} \hline \hline  
      index& $\zeta$ [Mm] & $\Phi$ [Mx] & T &      W      &  $H_m\,\,[{\mbox {Mx}}^2]$ \\ 
     ~ 0    &  0.79 &  -3.05e+18 &   -0.0138 &   ~ 0.0258 & ~ 1.11e+35\\
     ~ 4    &  0.80 &  -1.56e+18 &   -0.0272 &   ~ 0.0182 &  -2.19e+34\\
     ~ 9    &  1.06 &  -4.71e+18 &   -0.0210 &   ~ 0.0024 &  -4.12e+35\\
      15    &  0.92 &  -2.55e+18 &   -0.0224 &   -0.0163  &  -2.51e+35\\
      20    &  0.80 &  -5.69e+18 & ~  0.0111 &   ~ 0.0041 &  ~ 4.94e+35\\
      23    &  1.02 &  -2.64e+17 &   -0.0054 &   ~ 0.0166 &  ~ 7.80e+32\\
      25    &  1.00 &  -5.13e+17 & ~  0.0096 &   ~ 0.0079 &  ~ 4.59e+33\\
\hline 
      28   &  0.32 &  -4.43e+18 &   -0.0023 &   ~ 0.0000 & -4.42e+34\\
      29   &  0.33 &  -3.41e+18 &   -0.0028 &   ~ 0.0021 & -8.34e+33\\
      30   &  0.34 &  -6.29e+18 &   -0.0041 &   ~ 0.0011 & -1.20e+35\\
      36   &  0.34 &  -7.73e+18 &   -0.0065 &   ~ 0.0010 & -3.30e+35\\
      40   &  0.37 &  -6.30e+18 &   -0.0041 &   ~ 0.0015 & -1.01e+35\\
      41   &  0.36 &  -1.13e+19 &   -0.0001 &   ~ 0.0007 & ~ 7.56e+34\\
      46   &  0.35 &  -2.04e+19 &   -0.0010 &    -0.0017 &  -1.15e+36\\
\hline
    \end{tabular}
  \end{center}
\end{table}

In Table~\ref{table2} we present the resulting values of the axis length, magnetic flux, 
twist, writhe and self magnetic helicity for the flux tubes of the 14 largest zones. As the 
magnetic helicity depends on the square of the magnetic flux, and our selected areas are
very different in area, the resulting magnetic helicity varies over a wide range of several 
orders of magnitude. To study the dependence of the precedent quantities on the length of 
the respective axis, in 
Fig.~\ref{Fig3}, we plot the length of the axis of each flux tube $\zeta$, the writhe, 
the twist, and the sum of twist and writhe for the $54$ selected tubes related to intraspines 
(index ranging form $0$ to $26$) and spines (index ranging form $27$ to $53$). The values corresponding 
to tubes of the same zone (see left panel of Fig.~\ref{Fig1}) are connected by straight lines. 
As the magnetic field in the spines is more vertical than in the intraspines, the length of the axis 
of the tubes related to the 
spines is clearly shorter. For each intraspine/spine zone the length of the axis grows with the index 
because each of the related tubes was chosen in the interior of the preceding one. The 
writhe of the intraspines does not follow a clear pattern, but most of the tubes have a positive 
writhe. The spines show a nearly zero writhe. The twist of the intraspines, however, takes 
nearly always negative values. Consequently,
the twist is partially canceled by the writhe, thus the absolute value of the self magnetic helicity is,
 nearly always, lower than the absolute value of the twist.

In panels (a), and (b) of Fig.~\ref{Fig4} we plot the twist, and writhe per unit 
length versus the length of the axis of each flux tube $\zeta$. The normalized twist does 
not show a clear dependence with the length of the axis, but the absolute value of the 
normalized writhe in the intraspines clearly decreases with increasing $\zeta$. The axes of 
tubes related to intraspines are less wrung when the tubes are more horizontal 
(i.e., in the central part of the intraspines). As the writhe of the 
spines is very small, we can conclude that the writhe reaches only significant values 
when the tube includes the border of a intraspine. 
In panel (c) we plot the wavenumber (1/$pitch$), i.e., the number of turns done by the magnetic 
field per length unit as a function of $\zeta$. Flux tubes related to spines show slightly smaller 
values but any clear dependence is not observed.

In panel (d) we plot the normalized twist versus $\alpha/4\pi$ (i.e., the local twist 
$T^{\mbox{\rm\tiny{loc}}}$ at each pixel) evaluated at $z=200\,$km and averaged over 
each structure.
Given the good correlation between both magnitudes we can use the average of the local 
twist as a good proxy for the normalized twist. Consequently, we can explain the small 
obtained twist values in terms of the local twist: as we have seen in Fig.~\ref{Fig2}, 
$T^{\mbox{\rm\tiny{loc}}}$ reveals significant values with opposite sign at the borders 
of the penumbral filaments, leading to a cancellation of the twist when integrating over 
the filament. 

In order to assess the reliability of the most used proxies we 
plot, in panels (e) and (f), the average twist of each structure versus the average 
value of $\alpha_z/4\pi$ and $h_{c_{z}}$, respectively. Both proxies are very bad indicators 
of the average twist of intraspines but they give a qualitatively good 
agreement (better in the case of $h_{c_{z}}$) for the spines. This asymmetry
could be explained by the fact that both $\alpha_z$ and $h_{c_{z}}$
are only related with 
the vertical component $J_{z}$ of the electric current density vector,
which, as we show in Paper II, is much smaller than the horizontal 
components, mainly in and around the intraspines. On the other hand,
in a force-free configuration $\alpha_z=\alpha$ and $h_{c_{z}}=h_c$,
and then, the discrepancy between these parameters is a clear result of
the non-validity of the force-free approximation in the inner penumbra: 
In Paper II we have already shown that, at the borders of bright penumbral 
filaments, the magnetic field strongly departures from a force-free configuration.

\section{Conclusions}
In the present work we calculated the parameter $\alpha$ and its proxy $\alpha_z$, the 
current helicity density $h_{c}$ and its proxy $h_{c_{z}}$, the twist, the writhe, and 
the magnetic helicity of different structures of the inner penumbra of a sunspot. 
The parameters are evaluated from a three-dimensional geometrical model obtained after
the application of a genetic algorithm on inversions of spectropolarimetric data observed 
with {\it Hinode} \citep[see][Paper I]{puschmannetal10a}. We demonstrate, that in the inner penumbra the frequently 
used proxies $\alpha_z$ and $h_{c_{z}}$ are only qualitative indicators of the local twist 
(twist per unit length, evaluated under the assumption that the axis of a flux tube is 
parallel to the magnetic field) of penumbral structures. As shown in \citep[][Paper II]{puschmannetal10b}, the 
magnetic field in the area under study many times departs significantly from a force-free configuration 
and the horizontal component of the electrical current density is significantly larger than the vertical one.

The local twist shows only significant values at the borders of bright penumbral filaments and reveals 
opposite sign at each side of the bright filaments. The opposite sign might be the reason 
for a cancellation of the twist when integrating over the filament, thus the twist of 
the penumbral structures is very small. Significant values of the local twist are exactly 
related to areas where the electric current density is large. The local twist could be 
measuring the twist of the background field wrapping around the intraspines and/or the 
twist of the field lines of the intraspines themselves; in the latter case the internal structure 
of the tube would consist in two "cotyledons" (at both sides of a vertical plane 
longitudinally cutting the tube), harboring each one a magnetic field of opposite 
twist, compatible with the MHS model of \citet{borrero07}. 

The writhe per unit length diminishes with increasing length (decreasing inclination) of the 
axis of flux tubes related the intraspines. The small amount of twist and writhe shown
by the spines indicates that the background field lines, in these zones, are nearly straight.

A future study should clarify if the helicity apparent in the intensity maps of penumbral 
filaments in the mid and outer penumbra of sunspots is produced by helical flux tubes 
with a strong writhe or just by spurious effects produced by lateral intensity fluctuations. 
In any case, it is clear that a twisted tube does not {\it per se} generate  any intensity 
fluctuation similar to the observations by \citet{ryutovaetal08}. Rather, there is the 
necessity of a writhe of the tube, in such a way that different longitudinal portions 
of the tube were at different optical depths producing changes in the observed intensity. 
We will extend the present work (and necessarily the work presented in Paper I and II) 
on the entire sunspot.

\acknowledgments
We thank the referee for fruitful comments.
Hinode is a Japanese mission developed and launched by ISAS/JAXA, collaborating with 
NAOJ as a domestic partner, NASA and STFC (UK) as international partners. Scientific 
operation of the Hinode mission is conducted by the Hinode science team organized at 
ISAS/JAXA. This team mainly consists of scientists from institutes in the partner 
countries. Support for the post-launch operation is provided by JAXA and NAOJ (Japan), 
STFC (U.K.), NASA, ESA, and NSC (Norway).
 Financial support by the Spanish Ministry of Science and Innovation
through projects AYA2010--18029, ESP 2006-13030-C06-01, AYA2007-65602, and the European 
Commission through the SOLAIRE Network (MTRN-CT-2006-035484) is gratefully acknowledged. 
The National Solar Observatory (NSO) is operated by the Association of Universities
for research in Astronomy (AURA), Inc., under a cooperative agreement with the National 
Science Foundation. We thank V. Mart\'inez Pillet, C. Beck, and H. Balthasar for fruitful 
discussions.

\clearpage

\end{document}